# Influence of nanostructuring through high-pressure torsion (HPT) on superconductivity of a high-entropy alloy

Kaveh Edalati[1,2,]*, Alexy Bertrand[3], Payam Edalati[4], Thanh Tam Nguyen[1,2], Nariman Enikeev[5,6], Masaki Mito[3]

[1] WPI, International Institute for Carbon Neutral Energy Research (WPI-I2CNER), Kyushu University, Fukuoka 819-0395, Japan
[2] Mitsui Chemicals, Inc. - Carbon Neutral Research Center (MCI-CNRC), Kyushu University, Fukuoka 819-0395, Japan
[3] Graduate School of Engineering, Kyushu Institute of Technology, Kitakyushu 804-8550, Japan
[4] Faculdade de Engenharia Mecânica (FEM), Universidade Estadual de Campinas (UNICAMP), Limeira, Brazil
[5] Laboratory of Metals and Alloys under Extreme Impacts, Ufa University of Science and Technology, Ufa 450076, Russia
[6] Saint Petersburg State Marine Technical University, St. Petersburg 190121, Russia

High-entropy alloys (HEAs) have emerged as favorable choices for different applications, including superconductors. The present work examines the impact of nanostructuring via high-pressure torsion (HPT) on the superconducting properties of the equiatomic TiZrHfNbTa HEA. Structural characterization reveals a progressive refinement of grain size and increased dislocation density, together with partial phase transformation to an ω phase with HPT processing. Magnetic susceptibility and magnetization measurements indicate a systematic enhancement in the superconducting transition temperature (from $T_c$ = 6.2 K to 7.2 K) and critical magnetic field, as well as the stabilization of the superconductivity state by HPT processing. The improvement of superconducting properties is attributed to microstructural modifications such as grain boundary density, defect generation and phase transformations, and their impact on vortex pinning, quantum confinement and electron scattering. The results suggest that nanostructuring through severe plastic deformation provides an appropriate route to optimize superconducting properties in high-entropy superconductors.
**Keywords**: nanostructured materials; multi-principal element alloys (MPEAs); magnetic properties; dislocation density; ultrafine grain (UFG)

* Corresponding author: K. Edalati (E-mail: kaveh.edalati@kyudai.jp; Tel: +81-92-802-6744)

Superconductivity, characterized by zero electrical resistivity and the exclusion of magnetic field at temperatures lower than a critical temperature ($T_c$), is a quantum mechanical phenomenon with significant applications in magnetic resonance imaging, quantum computing and energy transmittance [1-3]. The performance of superconductors is strongly influenced not only by crystal structure and crystallinity but also by microstructure and nanostructure, making it crucial to explore how different structural and microstructural modifications affect superconducting properties [4-6]. Currently, the most popular research in superconductivity is the discovery of new superconductors with appropriate crystal and electronic structures; however, the ability to control microstructural and nanostructural features through advanced material processing techniques is considered essential for the development of next-generation superconducting devices [1-6].

High-entropy alloys (HEAs), with a minimum of five major elements [7-9], have currently gained attention as promising superconductors due to their unique electronic and lattice structures [10-15]. These multi-element alloys illustrate diverse properties like enhanced strength, oxidation resistance and tunable electronic structures, making them potential candidates for advanced superconducting applications [10-15]. The alloys $Ti_{11}Zr_{14}Hf_8Nb_{33}Ta_{34}$ ($T_c$ = 7.3 K) [16], $Ti_{1/6}Zr_{1/6}Hf_{1/6}Nb_{2/6}Ta_{1/6}$ ($T_c$ = 7.8 K) [17], $(TiZrHf)_{1/3}(NbTa)_{2/3}$ ($T_c$ = 7.7 K) [18], $Ti_{15}Zr_{24}Hf_{21}V_{15}Nb_{25}$ ($T_c$ = 5.3 K) [19], $Ti_{0.11}Zr_{0.11}Hf_{0.11}Nb_{0.11}Re_{0.56}$ ($T_c$ = 4.4 K) [20], $(NbTa)_{2/3}(HfMoW)_{1/3}$ ($T_c$ = 7.3 K) [21], $(NbTa)_{2/3}(MoWTh)_{1/3}$ ($T_c$ = 4.3 K) [22], $(ZrNb)_{1-x}(MoReRu)_x$ ($T_c$ = 3.4-5.3 K) [23], $(HfTaWIr)_{1-x}Re_x$ ($T_c$ = 2.1-5.9 K) [23], $(HfTaWPt)_{1-x}Re_x$ ($T_c$ = 7.3 K) [23] and $AgInSnPbBiTe_5$ ($T_c$ = 2.4-6.3 K) [24] are some examples of HEAs that show low-temperature superconducting behavior. Despite recent interests in high-entropy superconductors [10-15], the influence of microstructural and nanostructural features like grain refinement and defect generation on their superconducting properties remains underexplored in available publications [16-24].

Nanostructuring through severe plastic deformation processes, such as high-pressure torsion (HPT) [25,26], has been extensively employed to enhance the mechanical and functional properties of different metals, alloys and compounds [27,28]. HPT induces significant grain refinement, increases dislocation density and promotes phase transformations [25-28]. These changes directly impact superconducting properties by modifying vortex pinning, enhancing carrier scattering and altering electron-phonon coupling [29-31]. Increased grain boundary density provides additional pinning centers, which can strengthen superconducting coherence and inter-grain coupling [32-34]. While dislocations and relevant lattice strain are considered unfavorable for crystallinity and superconductivity, dislocations may serve as favorable sites for Cooper pair interactions [35,36]. The ability to control these microstructural features through HPT offers a powerful tool for tailoring the superconducting properties of metals and alloys [29-31]; however, no research has yet explored this aspect for high-entropy superconductors.

The present investigation aims to clarify the impact of HPT-induced nanostructuring on the superconducting performance of the equiatomic high-entropy alloy (HEA) TiZrHfNbTa, having a body-centered cubic (BCC) crystal form. It is found that the selected alloy, originally introduced by Senkov *et al.* for mechanical applications [37], exhibits superconductivity, with nanostructuring via HPT enhancing both the superconducting upper critical magnetic field and transition temperature. By analyzing microstructural modifications through X-ray diffraction (XRD), electron backscatter diffraction (EBSD) and transmission electron microscopy (TEM), and correlating these changes with superconducting measurements, the present work offers insights into the positive impact of grain refinement and defect generation on high-entropy superconductors. Moreover, the findings suggest a processing pathway using severe plastic deformation for developing advanced high-entropy superconducting materials with optimized performance.

The TiZrHfNbTa HEA was synthesized through arc melting using an atmosphere of highly pure argon (over 99.9999%) to avoid contamination by carbon, nitrogen, oxygen, etc. High-purity elements (over 99.7% purity) were melted in a copper crucible to make an ingot with about 10 g mass. The ingot was remelted seven times to obtain uniform elemental distribution, while the crucible was kept cool by circulating water. Discs measuring 5 mm in radius and 0.8 mm in height were cut from the final ingot by a wire-cutting machine, followed by homogenization at 1473 K for 1 hour and water quenching. These discs were then exposed to HPT processing at ambient temperature employing a high pressure of 6.0 GPa. The HPT process was performed for various revolutions ($N$ = 1/8, 1 and 10) to investigate the evolution of superconducting properties compared to the homogenized condition ($N$ = 0).

Structural characterization was performed using XRD to evaluate phase composition and lattice strain. XRD measurements with a scanning step of 0.05° per step and a scanning speed of 2° per minute were conducted using copper K$\alpha$ radiation. Dislocation density and crystallite size were obtained from XRD profiles using full-profile analysis with the Rietveld refinement method, implemented in the MAUD software [38]. The analysis followed the recommended approach for severely strained materials, with a focus on cubic crystal structures, as described in detail elsewhere [39]. EBSD was utilized for grain orientation and size analyses of the homogenized sample. EBSD maps were acquired under a voltage of 15 kV in a scanning electron microscope employing a 1 µm step size. The EBSD results were processed by the MTEX toolbox of MATLAB, in which regions with sizes below 3 pixels were included in the neighboring big grains, and boundaries of grains were smoothed by allocating the mean orientation of neighboring regions to non-indexed pixels by employing the meanOrientation function of the MTEX toolbox [40]. Transmission electron microscopy (TEM) was utilized to directly image grain refinement, dislocation structures and phase transformations. Disc samples with 1.5 mm radius were prepared from the edge of HPT-deformed samples and polished to 100 µm thickness, and later electrochemically polished employing a twin-jet electropolishing system at 463 K with 20 V using a chemical solution containing 65% methanol, 30% butanol and 3% perchloric acid (vol%). The last polishing to achieve thin foils was done by ion milling for 30 min employing argon at a voltage of 5 kV with ±5º polishing angles. TEM sample observations were carried out under a voltage of 200 kV to take conventional and high-resolution micrographs and achieve selected area electron diffraction (SAED) profiles as well as fast Fourier transform (FFT) patterns.

The superconductivity was studied by examining the magnetic shielding signal in magnetization employing a superconductivity quantum interference device (SQUID) with both alternating current (AC) and direct current (DC) measurement capabilities. A disc with a 1.5 mm radius and 0.5 mm height (nearly 30 mg) was prepared from the edge of HPT-processed samples and placed in the magnetometer so that the normal vector of the disc was placed perpendicular to the direction of AC and DC magnetic fields, ensuring that the diamagnetic field effect was not considered. The frequency of the AC magnetic field was 10 Hz and its amplitude was 0.4 mT (4 Oe). Magnetic measurements were conducted over a temperature range of 1.8-20 K to determine the superconducting transition temperature ($T_c$). For AC magnetization tests, the magnetization ($M$) data were recorded versus time ($t$), and the in-phase component of magnetization ($m'$) was evaluated by the Fourier analysis of the $M(t)$. The stability of superconductivity in a magnetic field and upper critical magnetic field ($H_{c2}$) were investigated by the DC magnetization measurements at different magnetic fields of 0.01, 0.5, 1.5, 2.0, 2.5, 3.0, 3.5, 4.0, 4.5 and 5.0 T.

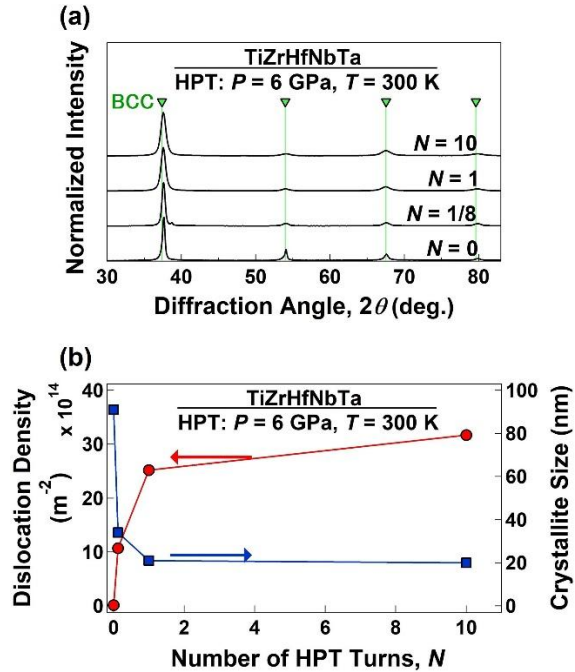

Figure 1. Increasing dislocation density and decreasing crystallite size by severe plastic deformation of high-entropy alloy. (a) XRD patterns, and (b) dislocation density and crystallite size estimated from the XRD patterns for TiZrHfNbTa treated via HPT for $N = 0$ (homogenization), 1/8, 1 and 10 revolutions.

XRD patterns in Fig. 1a indicate the presence of the BCC phase with no clear evidence for phase transition after HPT treatment. Calculations of lattice parameters using the Rietveld method indicate that the homogenized material and the sample treated via HPT for $N = 1/8$ have a lattice parameter of 3.4089 Å, while the lattice parameter slightly expands by 0.0007 Å with an increase in HPT turns. These small lattice expansions by HPT processing, if genuine, may be attributed to the redistribution of elements in the solid solution and/or the formation of vacancies, defects that have been reported in various HPT-processed metals [41,42] and ceramics [43,44]. Moreover, peak broadening occurs with HPT processing, signifying a reduction in crystallite size and an enhancement in lattice strain. The broadening (i.e. full width at half maximum of peaks) rises progressively with the number of HPT revolutions, indicating a larger induced strain and a higher density of lattice defects, like dislocations, and a decrease in crystallite size, as quantitatively shown in Fig. 1c. These results are a natural consequence of severe plastic deformation applied by HPT [25-29].

EBSD map for the homogenized samples in Fig. 2a and TEM analysis prior and after HPT treatment in Fig. 2b-e reveal a transition from coarse grains (>100 μm) in the homogenized sample to ultrafine grains (~20 nm) in the disc treated via HPT for $N = 10$. The grain refinement is accompanied by a significant enhancement of the fraction of high-angle grain boundaries, as evident from the ring patterns of SAED analysis. Such grain boundaries are supposed to function as effective pinning centers for magnetic flux vortices in superconductors [32-34]. It should be noted that some twin-like structures were detected after $N = 1/8$ as shown in Fig. 2b, which correspond to an ω phase. This phase was reported after HPT processing of titanium, zirconium and their alloys, although it cannot be recognized easily from the BCC phase using the XRD analysis [45-47].

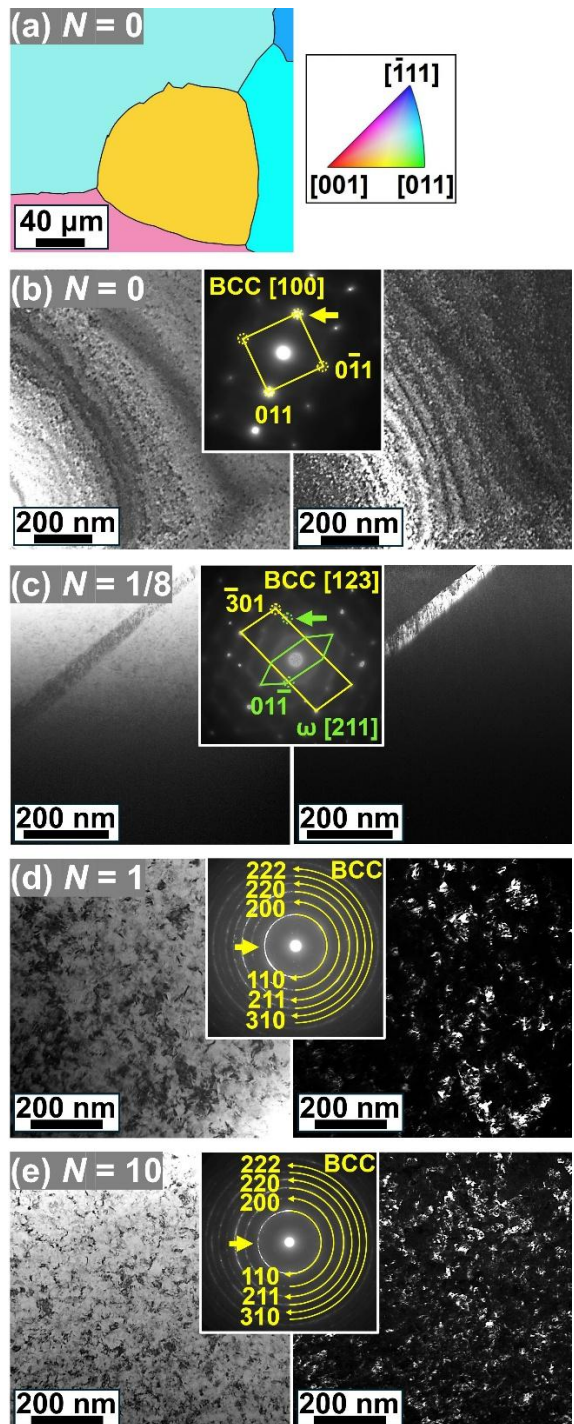

Figure 2. Transition from coarse grains to nanostructured grains by severe plastic deformation of high-entropy alloy. (a) EBSD crystal orientation mapping and (b-e) TEM bright-field micrograph (left side), SAED profile (center) and dark-field micrograph (right side) for TiZrHfNbTa treated via HPT for (a, b) $N = 0$ (homogenization), (c) $N = 1/8$, (d) $N = 1$ and (e) $N = 10$ revolutions. Dark-field micrographs were achieved using beam spots shown by arrows in SEAD profiles.

TEM high-resolution images in Fig. 3 indicate several important points regarding the nanostructural features of the sample processed under extreme HPT conditions using $N = 10$ turns. First, the presence of dislocations in the low-angle grain boundary form is confirmed in Fig. 3a, as shown more clearly using FFT analysis in Fig. 3b and a magnified lattice image in Fig. 3c. Second, individual dislocations are visible within the core of grains as shown in Fig. 3d and 3e. The presence of dislocations within the nanostructure is consistent with the Rietveld analysis in Fig. 1b. These dislocations are expected to influence electron scattering, thereby affecting superconducting properties [35,36]. Additionally, TEM analysis reveals the presence of nanoscale ω-phase precipitates in the heavily deformed sample, as shown in Fig. 3f using a lattice image and in Fig. 3g using FFT analysis. It should be noted that the detection of this phase in the BCC phase using XRD is typically challenging, but small differences in the FFT can be used for its identification, as illustrated in Fig. 3h and 3i. These precipitates may contribute to the enhancement of superconducting properties by providing additional pinning centers [1-6], as reported in some other superconductors such as Nb-Ti alloys [29].

AC magnetization measurements versus temperature, illustrated in Fig. 4a and 4b in two different magnifications, show the superconducting behavior of TiZrHfNbTa, caused by transition metals, with a progressive increase in the transition temperature with HPT processing. The changes in the transition temperature are shown quantitatively in Fig. 4c, indicating an increase from 6.2 K in the homogenized sample to 7.2 K for the disc treated via HPT for $N = 10$. The enhanced transition temperature suggests that increased grain boundary density and dislocation structures participate in improved superconducting coherence despite a reduction in crystallinity [29-36]. Such an increase in the transition temperature, previously reported in pure metals [30-34], appears to be valid in multicomponent HEAs, despite their inherently strained structure due to the presence of multiple elements [7-9].

DC magnetization plots achieved at various magnetic fields are shown in Fig. 5a-d for the samples prior and after the HPT treatment. The transition temperature for superconductivity decreases with increasing applied magnetic field for all samples, but the intensity of the diamagnetic signal remains significantly higher with increasing magnetic field for discs treated with $N = 1$ and 10. A summary of DC magnetization measurements is shown in Fig. 5e as a plot of superconducting upper critical magnetic field ($H_{c2}$) versus transition temperature for superconductivity. An enhancement in the upper critical magnetic field with deformation by HPT is observed, indicating stronger flux pinning and enhanced critical current capacity after HPT processing. By linear extrapolating the data in Fig. 5e to determine the critical magnetic field for disappearing superconductivity at 0 K, it is observed that the critical magnetic field for disappearing superconductivity increases from 14.2 T for the homogenized sample to 17.5 T for the disc treated via HPT for $N = 10$. Here, it should be noted that the behavior of the current high-entropy alloy at low fields does not follow the expected change in transition temperature from the Ginzburg-Landau (GL) equation, but it is still possible to achieve a satisfying fit by using some assumptions in the Werthamer-Helfand-Hohenberg (WHH) model. To do this, the critical field at $T = 0$ K was first calculated using the WHH model [14,48].

$$H_{c2}(T=0) = -0.693 \times T_c(H=0) \times dH_{c2}/dT(T=T_c) \tag{1}$$

where $T_c(H=0)$ is the transition temperature with zero magnetic field and $dH_{c2}/dT(T=T_c)$ is the derivative of the $H_{c2}$-$T_c$ curve near $H_{c2} = 0$. Because of the nonconventional behavior of high-entropy samples at low field, $T_c(H=0)$ and $dH_c/dT(T=T_c)$ values were selected, ignoring the low-field data. Then the $H_{c2}(T)$ data were fitted using an empirical equation [49].

$$H_{c2}(T) = H_{c2}(T=0)\left[1 - (T/T_c(H=0))^\beta\right] \tag{2}$$

where $\beta$ is a fitting parameter, which is adjusted to achieve the best fit. The fittings, obtained using this method, are given in Fig. 5f, indicating that the critical magnetic field for disappearing superconductivity follows the same trends as those achieved using linear extrapolation, although the estimated values are slightly lower. Data presented in Fig. 5e and 5f indicate that the superconducting state is thermodynamically stabilized by HPT processing. This stabilization is likely because of the influence of grain boundaries and dislocations on Cooper pair interactions, as expected from the Bardeen-Cooper-Schrieffer theory [50].

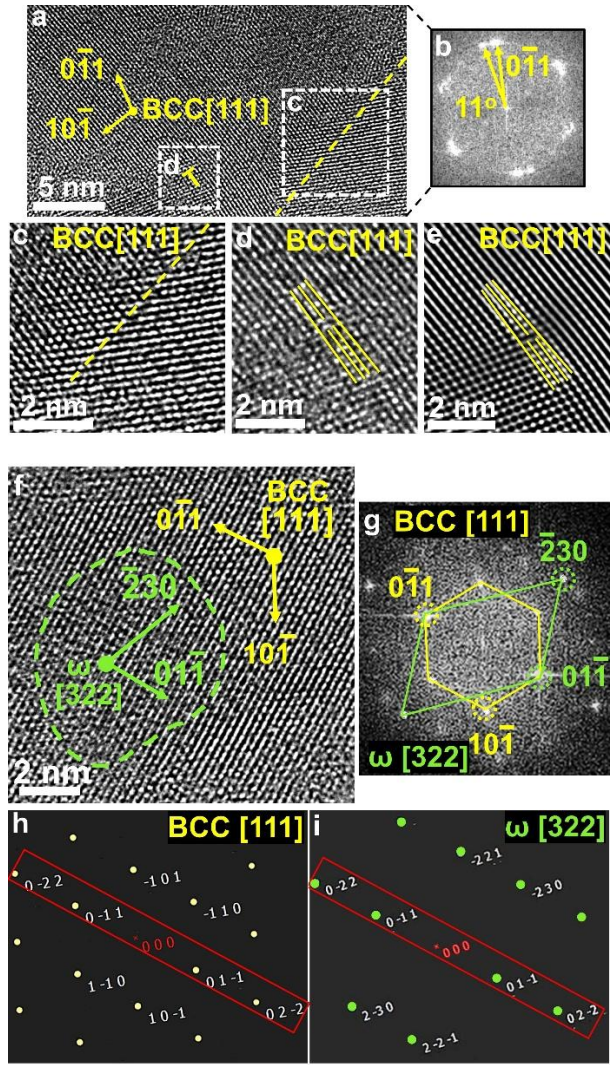

Figure 3. Formation of dislocations and ω phase by severe plastic deformation of high-entropy alloy. (a) High-resolution image, (b) FFT diffractograms taken from (a), lattice image of low-angle grain boundary from area shown using square in (a), (c) lattice image of a dislocation from area shown using square in (a), (d) inverse FFT analysis of lattice image of dislocation, (e) high-resolution image of region containing ω phase, (f) FFT diffractograms taken from (e), (g) simulated diffractogram of BCC phase and (h) simulated diffractograms of ω phase for TiZrHfNbTa treated via HPT for $N$ = 10 revolutions.

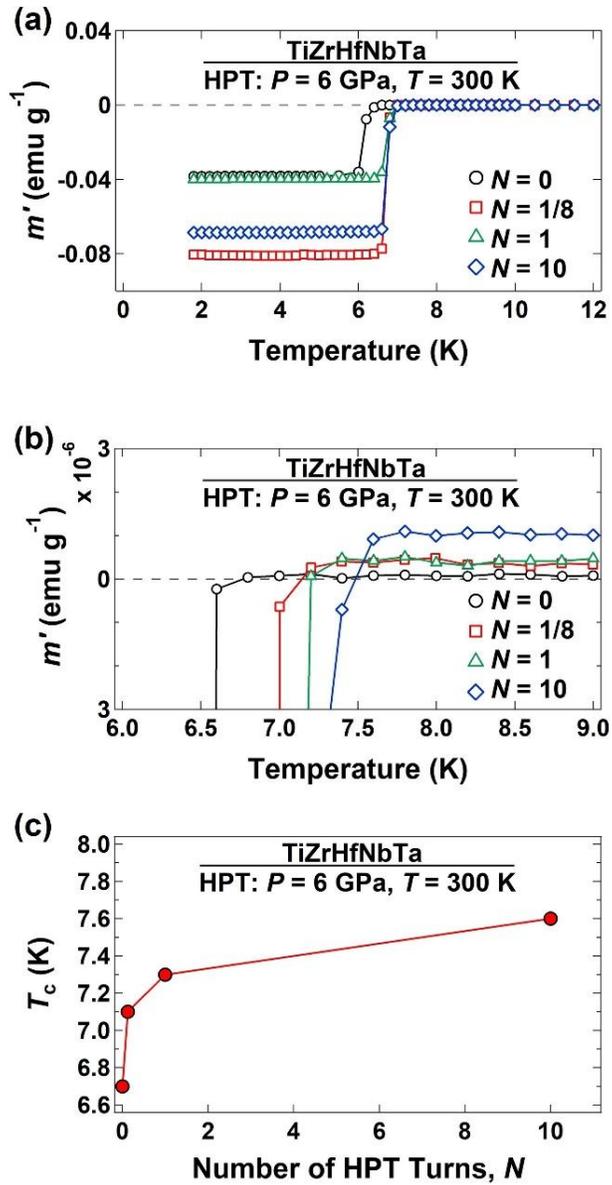

Figure 4. Enhancement of transition temperature for superconductivity by severe plastic deformation of high-entropy alloy. (a) Variations of in-phase AC magnetization ($m'$) versus temperature, (b) magnified view of (a) in superconducting transition region and (c) variations of critical temperature ($T_c$) for superconductivity versus number of HPT revolutions for TiZrHfNbTa treated via HPT for $N = 0$ (homogenization), 1/8, 1 and 10 revolutions.

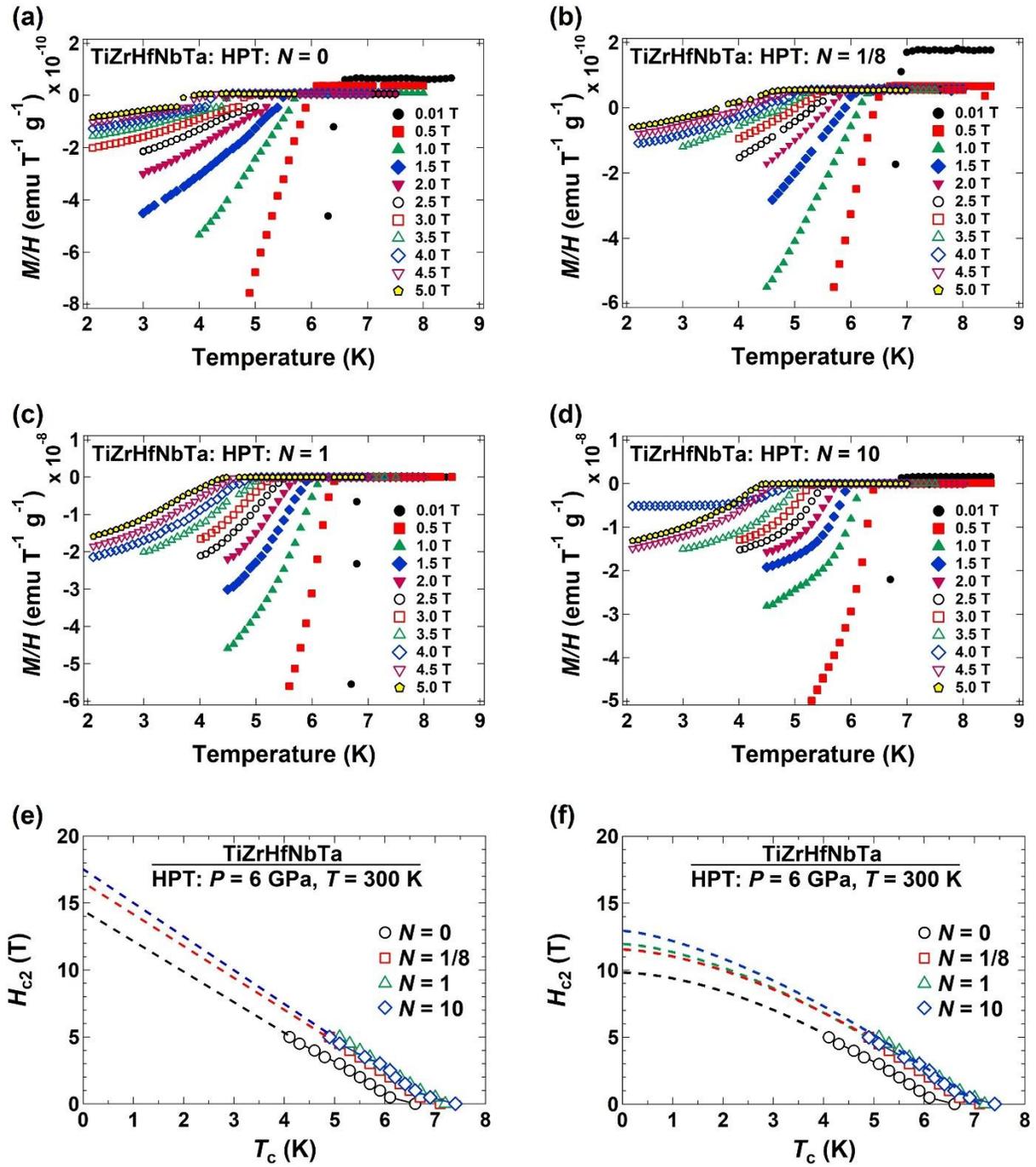

Figure 5. Increasing superconducting upper critical magnetic field by severe plastic deformation of high-entropy alloy. (a-d) Variations of magnetization ($M$) normalized by applied magnetic field ($H$) versus temperature at different applied magnetic fields from 0.01 T to 5 T for TiZrHfNbTa treated via HPT for (a) $N = 0$ (homogenization), (b) $N = 1/8$, (c) $N = 1$ and (d) $N = 10$ revolutions. (e, f) Variations of superconducting upper critical magnetic field ($H_{c2}$) versus transition temperature for superconductivity ($T_c$) achieved using data presented in (a-d), where the data in (e) and (f) were fitted using linear extrapolation and Werthamer-Helfand-Hohenberg model, respectively.

The current results not only introduce TiZrHfNbTa as a superconductor, expanding the number of reported high-entropy superconductors [10-24], but also demonstrate that nanostructuring through the HPT process thermodynamically stabilizes the superconducting state. Here, the observed increases in the critical temperature for superconductivity as well as the critical magnetic field for disappearing superconductivity by HPT processing warrant further discussion.

The increase in critical temperature with nanostructuring is attributed to enhanced electron-phonon interactions facilitated by nanograins and dislocations [50]. The high density of grain boundaries in nanostructured HEAs serves as effective pinning centers, reducing flux motion and improving superconducting phase coherence [32-34]. The role of dislocations in modifying superconducting behavior is twofold. While they act as additional pinning centers, they also introduce local strain fields, which can alter the density of states close to the Fermi level, affecting Cooper pair formation [35,36,50]. The observed increase in dislocation density with HPT processing is consistent with the enhancement of superconducting properties, as dislocations provide additional sites for vortex pinning and electron-phonon interactions. In addition to grain boundaries and dislocation-type lattice defects, the presence of the ω phase in deformed samples, as indicated by TEM analysis, further suggests that phase transformations may contribute to superconducting enhancement by providing vortex pinning centers, as widely reported in the Nb-Ti alloys [29]. The ω phase, which is typically stable at high pressures, is stabilized at room temperature in the presence of high lattice defect density generated by HPT treatment [45-47]. Since transformations to high-pressure phases are usually accompanied by a rise in the density of states near the Fermi level [51,52], they can enhance electron-phonon coupling and contribute to observed increases in critical current and magnetic field for superconductivity [50]. Another explanation using the Bardeen-Cooper-Schrieffer theory for the enhancement of superconducting properties by HPT is the finite-size and quantum confinement effects. When grain sizes are reduced to the nanoscale, quantum confinement effects can enhance pairing interactions by modifying the phonon spectrum and increasing the superconducting gap, thereby raising the critical temperature and field [53,54].

Overall, the observed increase in critical temperature and magnetic field for superconductivity with increasing HPT revolutions highlights the potential of severe plastic deformation techniques for tailoring superconducting properties in HEAs. It should be mentioned that although HPT is a high-pressure technique, superconductivity after HPT is usually examined under ambient pressure [28-30]; whereas other high-pressure methods are employed to evaluate the superconductivity under high pressure [14,55-57]. Examination of superconducting properties of HPT-processed materials under high pressure, which has been limitedly investigated for conventional superconductors [58], can be a future research direction for high-entropy superconductors. Moreover, the expansion of this study to high-entropy ceramics, such as high-entropy oxides and oxynitrides [59] or even high-entropy hydrides [60], can be of future interest to understand their superconducting property changes by HPT processing.

In conclusion, this study demonstrates that HPT processing significantly enhances the superconducting properties of TiZrHfNbTa. The increased critical temperature and magnetic field observed with increasing HPT revolutions is attributed to refined grain size, increased dislocation density and partial phase transformation to a high-pressure phase. These findings provide a pathway to optimizing superconducting performance for technological applications by engineering grain boundaries and defect structures. Moreover, this study demonstrates that HPT processing is an efficient method to enhance the properties of high-entropy superconductors, paving the way for their use in advanced cryogenic applications.

**CRediT Authorship Contribution Statement**

All authors: Conceptualization, Methodology, Investigation, Validation, Writing – review & editing.

**Declaration of competing interest**

The authors declare no competing interests that could have influenced the results reported in the current article.

**Acknowledgements**

This research is supported partly by a Grant-in-Aid from the Japan Society for the Promotion of Science (JP22K18737), and partly by the Ministry of Science and Higher Education of the Russian Federation as part of the World-Class Research Center Program: Advanced Digital Technologies (contract No. 075-15-2022-312 dated 20 April 2022).

**Data Availability**

The data will be available to share upon request from the corresponding author.